\documentclass[]{article}
\begin{document}
\title{Static Cylindrical Matter Shells}
\author{Metin Ar\i k\footnote{{Department of Physics, Bogazici University,  Bebek, Istanbul, Turkey; e-mail: arikm@boun.edu.tr}}
\  and \"{O}zg\"{u}r Delice\footnote{Department of Physics,
Bogazici University,  Bebek, Istanbul, Turkey; e-mail:
odelice@boun.edu.tr}} \maketitle

\begin{abstract}  Static cylindrical shells composed of massive
particles arising from matching of two different Levi-Civita
space-times are studied for the shell satisfying either isotropic
or anisotropic equation of state. We find that these solutions
satisfy the energy conditions for certain ranges of the
parameters.
\end{abstract}



\maketitle
\section{Introduction}
The cylindrically symmetric static vacuum space-time is given by
the Levi-Civita metric \cite{levicivita}, which can be written in
the form
\begin{equation}
ds^{2}=-q^{4s }dt^{2}+ q^{4s (2s -1)}(dq ^{2}+ dz^{2}) +\alpha^2
q^{2(1-2s)}d\phi^{2} \label{Lc}
\end{equation}
where $t$,  $q$, $z$ and $\phi$  are the time, the radial,
 the axial, and the angular coordinates with
the ranges $-\infty <t< \infty$, $0 \leq q < \infty$,
$-\infty<z<\infty$ and $0<\phi \leq 2\pi$. This metric has two
real parameters which cannot be removed by a coordinate
transformation. Cylinders \cite{Bonnor, philbin} and cylindrical
shells \cite{Stachel}-\cite{gleiser} were studied as a physically
acceptable sources of this metric in order to understand  the
meaning and the behavior of these parameters. For recent reviews
on cylinders see \cite{Bonnor}. The first parameter, $s$, is
related to the energy density of the source whereas the second
one, $\alpha$, is related to the global conicity of space-time.
When $s=0$ or $1/2$ the space-time is flat and corresponds to
Minkowski space-time. For $0<s<1/2$ this metric is the exterior
metric of an infinite line mass (line singularity). For $s=0$ the
metric is flat and if $\alpha=1$ it is regular whereas if $\alpha
\neq 1$ it becomes the
 well known exterior cosmic string metric \cite{Vilenkin}. For $s=1/2$ this
metric correspond to an infinite plane \cite{dasilva}. In this
paper we limit our analysis to $-1/2\leq s \leq 1/2$.

Our aim is to study infinitely thin cylindrical shells composed of
matter particles having isotropic or anisotropic equation of state
and satisfying certain energy conditions as a source of exterior
Levi-Civita metric. Actually, it is known \cite{zouzou}-
\cite{gleiser},\cite{bicak} that the matching conditions yield the
conditions for the shell to be matter shell or shell composed of
massless particles. The shells composed of photons is studied
recently in detail. Some properties of the photonic shells around
a regular interior or a cosmic string \cite{bicak} or a line
singularity \cite{AD1} have been discussed and generalized to the
multiple shell case \cite{AD2}. The particle motion about a
photonic shell with flat interior and exterior is discussed in
\cite{Zofka}. In this paper, we will focus on  non-photonic
shells-matter shells- having isotropic or anisotropic equation of
state. Imposing several equation of states for the shell one can
find admissible values of the metric parameters satisfying some
energy conditions. We will compare the results with the previous
works on this topic, especially with \cite{bicak} which studied
shells made of various types of matter as a source of the
Levi-civita metric. They relate the interior properties of the
shell,especially the mass per unit lenght of the shell with the
two parameters of the exterior metric.  They limited their
analysis with the case where the interior region is flat or
locally flat and  a cosmic string can be present on the symmetry
axis. We also refer \cite{bicak} for a review of previous works on
this topic since a good comparison of their results with the
previous works has been given. Our main difference is that we will
allow the inside of the shell not flat. This may not seem
physically attractive, however this is not the case, since the
shell can surround another shell or cylinder satisfying (or not
satisfying) physical requirements. Specifically, the shell can be
one member of a multiple shell system\cite{gleiser,AD2}. Thus, the
interior region of the shell is not necessarily flat in the most
general case.

\section{Matching conditions}

We can write the Levi-Civita metric in normal form by transforming
the radius $q$ into a proper radius $r$ by defining $dr=q^{2s(2s
-1)}dq$
which results in 
\begin{equation}
ds^{2}=-R^{4s /N}dt^{2}+dr^{2}+ R^{4s (2s -1)/N}dz^{2} +\alpha^2
R^{2(1-2s )/N}d\phi^{2}, \label{Lc1}
\end{equation}
where
\begin{equation}\
 q =R^{1/N},\ \  R=Nr, \ \ N=4s ^{2}-2s +1.
\end{equation}

Let us denote the interior and the exterior radial coordinates as
$r_\pm$ and the location of the shell at $(r_{\pm}=r_{0\pm})$. We
want both the interior and the exterior metrics to be continuous
on the shell. After rescaling the metrics, the interior and the
exterior metrics can be written as:

\begin{eqnarray}
&&ds^2_-=-r_-^{\frac{4s}{N}}dt^2+dr_-^2+r_-^{\frac{4s(2s-1)}{N}}dz^2+\alpha_-^2
r_-^{\frac{2(1-2s)}{N}}d\phi^2,\quad (r_-\leq r_{0-}), \\
&&ds^2_+=-r_+^{\frac{4s'}{N'}}dt'^2+dr_+^2+r_+^{\frac{4s'(2s'-1)}{N'}}dz'^2+\alpha_+^2
r_+^{\frac{2(1-2s')}{N'}}d\phi^2,\ \ (r_+\geq r_{0+}).
\phantom{aa}
\end{eqnarray}

To achieve the continuity at $r_-=r_{0-}$, $r_+=r_{0+}$, we define
$t'=Kt$, $z'=Lz$, then we find:
\begin{equation}
K=\frac{r_{0-}^{2s/N}}{r_{0+}^{2s'/N'}}, \quad
L=\frac{r_{0-}^{2s(2s-1)/N}}{r_{0+}^{2s'(2s'-1)/N'}}, \quad
\alpha_+=\frac{\alpha_- r_{0-}^{(1-2s)/N}}{r_{0+}^{(1-2s')/N'}}.
\end{equation}

After a straightforward calculation using Israel's thin shell
formalism \cite{israel}, the
  Energy-Momentum tensor of the shell, whose form is
anisotropic in general with
$S^i_{\phantom{i}j}=diag\left(-\rho,p_z,p_{\phi} \right)$, can be
written as (we take $c=G=1$):

\begin{eqnarray}\label{emt}
-8\pi \rho&=&\frac{N'-2s'}{N'r_{0+}}-\frac{N-2s}{Nr_{0-}}
=8\pi(p_z+p_\phi)+\frac{4s}{Nr_{0-}}-\frac{4s'}{N'r_{0+}},\phantom{AAA}\\
8\pi
p_z&=&\frac{1}{N'r_{0+}}-\frac{1}{Nr_{0-}},
\quad \quad 8 \pi p_{\phi}=\frac{4s'^2}{N'r_{0+}}-\frac{4s^2}{Nr_{0-}}.\label{emtpp}
\end{eqnarray}

\section{Matter Shells}

 The trace of the energy momentum tensor of the shell
(\ref{emt}-\ref{emtpp}) is
\begin{equation}
4 \pi\, S=\frac{1}{r_{0+}}-\frac{1}{r_{0-}}.
\end{equation}

Thus, the vanishing of the trace of energy-momentum tensor of the
shell requires $r_{0+}=r_{0-}.$ Equivalently, if we perform a
coordinate transformation on  the inner or the outer independent
radial coordinates $r_-$ or $r_+$ to express one of them in terms
of the other, than the condition can be expressed as follows: If
there is no relative shift in the values of the interior and the
exterior radial coordinates on the shell, than the shell is
composed of massless particles. If there is a shift, than the
shell is composed of massive particles.

\section{Isotropic Shells}
In this section we study shells with $p_z=p_\phi=p.$ Also we
impose an equation of state $p=\gamma \rho.$ where $\gamma$ is a
constant. These constraints give
\begin{equation}\label{isotropicgamma}
r_{0+}=\frac{(1-4s^2)N'\,r_{0-}}{(1-4s'^2)N},\quad
\gamma=\frac{s+s'}{(2s-1)(2s'-1)}.
\end{equation}
From these equations, we can find the exterior metric parameters
for given $\gamma$ and interior metric parameters or vice versa.
The energy density and the pressure become:
\begin{equation}\label{isotropicpress}
\rho=\frac{(s-s')(2s-1)}{2 \pi(1+2s')N r_{0-}},\quad
p=\frac{s^2-s'^2}{2\pi(4s'^2-1)N r_{0-}}.
\end{equation}
The values of these quantities for flat interior can be found by
taking $s=0.$

\begin{figure}[h]
\input{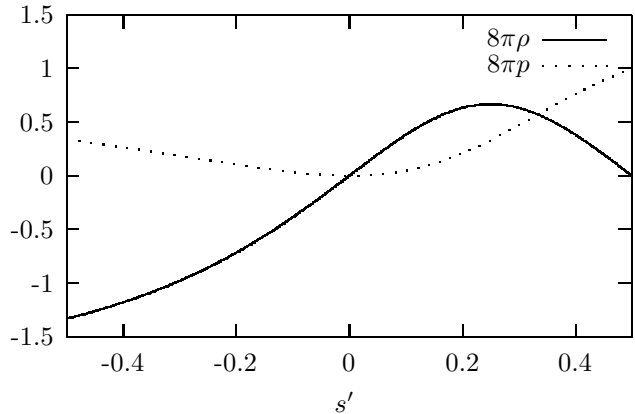}
\caption{The variation of the energy density and the pressure as a
function of the parameter $s'$ of the isotropic shell for the case
$s=0$ when $r_{0-}=1$.} \label{perfect}
\end{figure}

For a shell with the stress-energy tensor of the form
$S_{ij}=diag(\rho,p_z,p_\phi)$ we have the weak energy condition
(WEC) with $\rho\geq 0$, $\rho+p_z\geq 0$, $\rho+p_\phi \geq 0$;
the dominant energy condition (DEC) with $\rho \geq 0$, $\rho\geq
p_z\geq - \rho$, $\rho\geq p_\phi\geq - \rho$; and the strong
energy condition (SEC) with  $\rho+p_z+p_\phi \geq 0$ together
with the conditions for WEC \cite{HawkingEllis}. For $s=0$ this
shell satisfies WEC and SEC for $0\leq s' \leq 1/2$, DEC for
$0\leq s' \leq 1/3$. Notice that when the interior metric
parameter $s$ is getting bigger, the range of $s'$ when the shell
satisfies the energy conditions decreases. For $s=1/2$, $\rho$
vanishes.  And for $s>1/2$ there is no solution satisfying the
energy conditions except WEC since $\rho$ is always smaller than
$p$ for this case. These results are in agreement with the
previous work \cite{bicak} who studied this case for $s=0$. The
changing of the energy density and pressure with $s'$ is presented
in Figure \ref{perfect} for the case when the interior metric
parameter $s=0$. As previous studies of the cylindrical shells
suggest \cite{bicak}, the energy density of the shell is bounded
from above.

\subsection{Some examples on isotropic shells}
If we choose $\gamma=0$ in (\ref{isotropicgamma}) we find a
solution with shell composed of pressureless dust. We have
\begin{equation}
s'=-s,\quad r_{0+}=\frac{1+2s'+4s'^2}{N'}r_{0-}, \quad
\rho=\frac{s'}{\pi N r_{0-}}.
\end{equation}

\begin{figure}[h]
\input{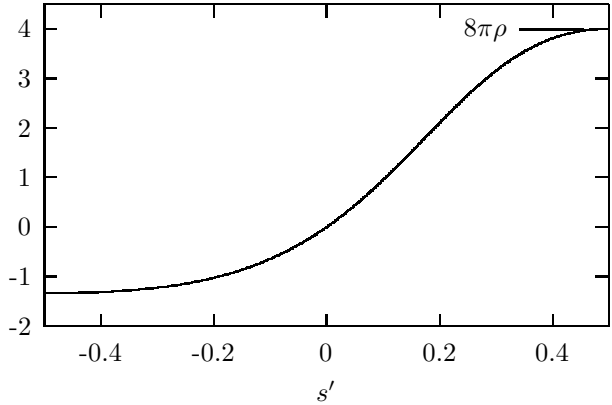}
\caption{The variation of the energy density of the pressureless
dust shell as a function of $s'(=-s)$ for $r_{0-}=1$.}
\label{dust}
\end{figure}

We know that Levi-Civita metrics with negative parameter are
isomorphic to the positive parameter cases and correspond to a
repulsive (negative energy) line mass. In this case, the energy
density of this shell  is positive when $s$ is negative and vice
versa. Thus , when the interior metric parameter $s$ is negative,
the shell satisfies all the energy conditions. The energy density
 of the shell is bounded from above and reaches its maximum value for $s'=1/2$. The
variation of the energy density for a given $s'$ is presented in
Figure \ref{dust}.

Domain walls are topological defects which might arise during the
phase transitions of the early universe \cite{Vilenkin}. They have
the equation of state $\gamma =-1$. Applying our shell to this
equation of state we find that,
\begin{equation}
s'=\frac{s-1}{4s-1},\quad r_{0+}=-r_{0-}\quad \rho=-p=\frac{1}{6
\pi r_{0-}}.
\end{equation}

 Thus, we can have a  domain wall with positive energy
density provided that the outer radius of the shell is negative of
the inner radius of the shell. Thus, this solution is not
physically attractive.

We can have a shell with isotropically moving photons in $z$ and
$\phi$ directions with $\gamma=1/2$. Then we find
$s'=(s-1)/(4s-1)$ and $r_{0+}=r_{0-}$, thus inner and outer radii
agree for this case as expected.

\section{Anisotropic shells}

Here we will consider the case where the shell is composed of
anisotropic massive particles having the equation of state
$p_z=\kappa\, \rho$, $p_\phi=\lambda\, \rho$ where $\kappa$ and
$\lambda$ are constants. From (\ref{emt}-\ref{emtpp}) we find
\begin{eqnarray}
\kappa=\frac{N r_{0-}-N' r_{0+}}{(1-2s)^2N'r_{0+}-(1-2s')^2 N
r_{0-}},\\
\lambda=\frac{4s'^2Nr_{0-}-4s^2N'r_{0+}}{(1-2s)^2N'r_{0+}-(1-2s'^2)N
r_{0-}}.
\end{eqnarray}

This result is general and for any given $\kappa$ and $\lambda$ we
can determine the outer metric parameters in terms of the inner
metric parameters or vice versa. Then using
(\ref{emt}-\ref{emtpp}) we can determine their energy density and
pressure. Let us present some examples.
\subsection{Some examples on anisotropic shells}

We  first present dust-like solutions where $\kappa+\lambda=0$.
Applying these constraints on the field equations, we find
\begin{eqnarray}
&&r_{0+}=\frac{(1+4s'^2)N r_{0-}}{(1+4s^2)N'},\quad
\rho=\frac{s'(1+4s^2)-s(1+4s'^2)}{2\pi(1+4s'^2) N\,r_{0-}},  \\
&& p_z=-p_\phi=\frac{(s^2-s'^2)}{2\pi(1+4s'^2) N r_{0-}}.
\end{eqnarray}

\begin{figure}[h]
\input{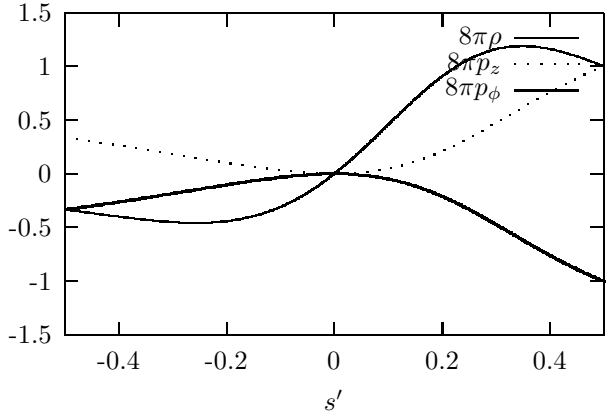}
\caption{The variation of the $\rho$, $p_z$ and $p_\phi$ of
dustlike shell as a function of $s'$ when $s=0$ and $r_{0-}=1$.}
\label{dustlike}
\end{figure}

The changing of the energy density and the pressures for the
interior metric parameter $s=0$ is presented in figure
\ref{dustlike} for $r_{0-}=1$. For this case the shell satisfies
WEC, DEC and SEC for $0\leq s'\leq 1/2$.

Next, let us discuss an anisotropic shell solution 
which
 corresponds to a hollow cylinder with infinitely long
straight cosmic strings \cite{Vilenkin}
with the equation of state $\rho=-p_z$.
 The exterior metric should  be the same of the exterior metric of
 infinitely long static cosmic string metric \cite{Vilenkin}. Indeed, only for $s'=0$ we have a solution of this
form with $s=0$ or $s=1$ which gives
\begin{equation}
8 \pi\rho=\frac{1}{r_{0-}}-\frac{1}{r_{0+}},\quad (s=0);\quad 8
\pi \rho= \frac{1}{3r_{0-}}-\frac{1}{r_{0+}},\quad (s=1).
\end{equation}

These solutions correspond to a cosmic string shell and in the
first one the interior region is flat \cite{zouzou,Tsoubelis}.
Same results are also presented in \cite{bicak}.

Let us now discuss a massive shell with only counter rotating
particles or counter moving particles in $z$ direction. For a
shell with counter rotating particles ($p_z=0$) we need
$\kappa=0$, then we find:
\begin{equation}
r_{0+}=\frac{Nr_{0-}}{N'},\quad  \rho=\frac{s'(1-s')-s(1-s)}{2\pi
Nr_{0-}},\quad  p_\phi=\frac{s'^2-s^2}{2\pi Nr_{0-}}.
\end{equation}
For flat interior, ($s=0$) this shell has positive energy density
for $0\le s'\le 1$, and satisfies WEC and DEC for this range
whereas  SEC for $0\le s'\le 1/2$.

For a shell with counter moving particles in $z$ direction
($p_\phi=0$) we need $\lambda=0$ which yields:
\begin{equation}
r_{0+}=\frac{s'^2Nr_{0-}}{s^2N'},\quad
\rho=\frac{s'^2(1-4s)-s^2(1-4s')}{8\pi s'^2Nr_{0-}},\quad
p_z=\frac{s^2-s'^2}{8\pi s'^2Nr_{0-}}.
\end{equation}
For this shell for flat interior there is no solution. For
negative $s$, we can find solutions satisfying all the energy
conditions, and for $0<s<1$ there is no solutions satisfying all
the energy conditions. Only for the interval $s > 1/2$ $\cap$ $s/
(4 s-1) < s' < s$ WEC and SEC is satisfied.

\section{Two-fluid shell}

It is interesting to note that the shell also admits a two-fluid
solution with $S_{ij}=S_{ij}^{(p)}+S_{ij}^{(m)}$ with
$S_{j}^{i(p)}=diag(-\rho^{(p)},p_z^{(p)},p_\phi^{(p)})$ and
$S_{j}^{i(m)}=diag(\rho^{(m)},p_z^{(m)},p_\phi^{(m)})$. We first
make a coordinate transformation to the outer radial coordinate as
$r_{+}=r_{-}+a$. Then we find:
\begin{eqnarray}
&&\rho^{(p)}=\frac{s'N-sN'}{4\pi NN'(r_{0-}+a)},\quad
p_z^{(p)}=\frac{N-N'}{8 \pi NN'(r_{0-}+a)} \\
&&p_{\phi}^{(p)}=\frac{s'^2N-s^2N'}{2 \pi NN'(r_{0-}+a)},\quad
\rho^{(m)}=\frac{a(N-2s)}{8\pi N r_{0-}(r_{0-}+a)},\\
&&p_{z}^{(m)}=\frac{-a}{8\pi N r_{0-}(r_{0-}+a)},\quad
p_{\phi}^{(m)}=\frac{-s^2a}{2\pi N r_{0-}(r_{0-}+a)}.
\end{eqnarray}
It is easy to see that $S_{ij}^{(p)}$ is traceless but
$S_{ij}^{(m)}$ is not. Thus, this solution represents a two-fluid
shell composed of a mixture of photons and matter. When $s=s'$ the
photonic part vanishes whereas  for $a=0$ the matter part
vanishes. Thus the existence of photons in the shell deviates the
outer metric parameter $s'$ from the inner metric parameter $s$
and the existence of matter particles results a shift in the
values of the inner and the outer radius of the shell. When $a=0$
we get photonic shells discussed recently in great detail
\cite{bicak,AD1}. For the matter part, the pressures of the shell
in $z$ and $\phi$ directions are negative. When the interior is
locally flat, the matter part satisfies $\rho=-p_z$ and this
two-fluid represents a shell  composed of both cosmic strings and
photons \cite{Wang,gleiser}.

\section{Conclusions}

In this work, we studied  some exact solutions of the Einstein
equations corresponding to static cylindrical shells composed of
massive particles. The shells constructed from matching of two
different Levi-Civita space-times can be classified as matter or
photonic shells whether their radial coordinates have relative
shifts or not. In this paper we have focused on to the matter
shells. The solutions we have presented can be seen as sources for
exterior Levi-Civita vacuum space-time. Our difference is that we
have chosen the interior space-time non-flat and  it is described
by a Levi-Civita metric and contains a line singularity at the
symmetry axis in general. However we can replace this singularity
with  a shell or a cylinder with finite radius.  Choosing the
interior not flat, we can consider the possibility that the shell
can be a part of a multiple shell system. The generalization of
these results to  multiple shell system is straightforward.
Choosing the interior of the innermost shell regular, we can avoid
having a singularity at the axis. Two concentric shells
\cite{gleiser} or multiple shells composed of photons \cite{AD2}
is presented recently. Thus considering previous works and our
results, one can find multiple shells having different equation of
states and satisfying certain energy conditions.

 We have imposed isotropic or anisotropic equations of state for the shell.
 Some solutions satisfy energy conditions.
 For example, the perfect fluid,
dustlike shell, shells with $p_z=0$, satisfies the weak, the
strong and the dominant energy conditions for some ranges of the
parameters. For a fixed $s$, one can find these ranges which are
only depend on $s'$ and the  interior and the exterior radial
coordinates on the shell. The shells can have any radius. Their
energy density and mass per unit length are bounded from above.
 For the pressureless dust shell, it satisfies all of the
energy conditions with the expense of negative line mass at the
axis. This is also true for a shell with particles moving only in
the axial direction. Similar behavior has been observed for a
static cylindrical shell with counter moving photons in $z$
direction \cite{AD1}. This might be an expected result since for
these cases there is no pressure in the fluid against collapse
($p_\phi=0$) and we need an repelling force to keep these shells
static.

\section*{Acknowledgements}
We thank Ahmet Baykal for reading the manuscript and help with the
figures.


\end{document}